\documentclass[superscriptaddress,twocolumn,showpacs,preprintnumbers,amsmath,amssymb,pra,longbibliography]{revtex4} 

\usepackage[dvipdfmx]{graphicx}
\usepackage{dcolumn,upgreek}
\usepackage{bm}
\usepackage{color}

\def\ii{{\mathrm{i}}}
\def\ff{{\mathrm{f}}}
\def\ee{{\mathrm{e}}}
\def\dd{{\mathrm{d}}}

\def\no{{\nonumber}} 

\def\bra#1{\langle #1|}

\def\ket#1{|#1\rangle}

\def\bracket#1{\langle #1 \rangle}
\def\bracketi#1#2{\langle #1 | #2 \rangle}
\def\bracketii#1#2#3{\langle #1 | #2| #3\rangle}

\def\sub#1{_\mathrm{#1}} 

\def\vct#1{{\mathchoice{\mbox{\boldmath$#1$}}{\mbox{\boldmath$#1$}}
  {\mbox{\scriptsize\boldmath$#1$}}{\mbox{\scriptsize\boldmath$#1$}}}}

\begin{document}


\title{Classical reconstruction of interference patterns of position--wavevector-entangled photon pairs by time-reversal method 
}

\author{Kazuhisa Ogawa}
\email{ogawak@ist.hokudai.ac.jp}
\affiliation{%
Graduate School of Information Science and Technology, Hokkaido University, Sapporo 060-0814, Japan
}%
 
\author{Hirokazu Kobayashi}
\affiliation{%
School of System Engineering, Kochi University of Technology, Tosayamada-cho, Kochi 782-8502, Japan
}%

\author{Akihisa Tomita}
\affiliation{%
Graduate School of Information Science and Technology, Hokkaido University, Sapporo 060-0814, Japan
}%

\date{\today}

\begin{abstract}
The quantum interference of entangled photons forms a key phenomenon underlying various quantum-optical technologies. 
It is known that the quantum interference patterns of entangled photon pairs can be reconstructed classically by the time-reversal method; however, the time-reversal method has been applied only to time--frequency-entangled two-photon systems in previous experiments.
Here, for the first time, we apply the time-reversal method to the position--wavevector-entangled two-photon systems: the two-photon Young interferometer and the two-photon beam focusing system.
We experimentally demonstrate that the time-reversed systems classically reconstruct the same interference patterns as the position--wavevector-entangled two-photon systems.


\end{abstract}
\maketitle



\section{Introduction}

Entangled photon pairs have been utilized for observing various quantum-optical phenomena that lie out of the scope of classical optics. 
In most experiments, entangled photon pairs are generated by spontaneous parametric down-conversion (SPDC).
The generated photon pairs can exhibit entanglement over several kinds of degrees of freedom such as time--frequency, position--wavevector \cite{aiueo}, polarization \cite{PhysRevLett.75.4337}, and orbital angular momentum \cite{mair2001entanglement}.
Depending on the kind of entanglement, they exhibit different quantum-optical phenomena.
For instance, time--frequency-entangled photon pairs have been used for observing automatic dispersion cancellation \cite{PhysRevA.45.6659,PhysRevLett.68.2421} in Hong--Ou--Mandel (HOM) interference \cite{PhysRevLett.59.2044}, Franson interference \cite{PhysRevA.45.3126,PhysRevLett.65.321}, and phase superresolution \cite{PhysRevLett.66.1142}.
In other instances, position--wavevector-entangled photon pairs have been used for observing ghost imaging \cite{PhysRevA.52.R3429}, two-photon Young interference \cite{PhysRevLett.82.2868,PhysRevLett.85.2733,PhysRevLett.87.013602,kawabe2007quantum,PhysRevLett.112.223602}, two-photon focused beam spots \cite{xu2015experimental}, and automatic aberration cancellation \cite{PhysRevLett.101.233603}.

On the other hand, recent studies have shown that two-photon detection patterns in an entangled two-photon system can be reconstructed classically by use of its time-reversed system.
This method, which is called \textit{the time-reversal method}, is based on the time-reversal symmetry of quantum mechanics: projection probabilities in the time-reversed system are equal to those in the time-forward system \cite{pregnell2005retrodictive,PhysRevLett.98.223601}.
Interestingly, with some ingenuity, the time-reversed system of a two-photon system can be prepared by the use of completely classical optical systems, with such a system including a classical light source, optical intensity measurement, and nonlinear optical transform.
In fact, the classical reconstruction of two-photon detection patterns via the time-reversal method has been experimentally demonstrated for certain quantum-optical phenomena caused by time--frequency-entangled photon pairs, such as automatic dispersion cancellation in HOM interference \cite{kaltenbaek2008quantum,PhysRevA.91.013846}, dispersion-cancelled OCT \cite{Lavoie2009,mazurek2013dispersion,ogawa2016classical}, and phase superresolution \cite{PhysRevLett.102.243601,PhysRevA.88.063813}.

These previous studies have been conducted via the application of the time-reversal method to time--frequency-entangled two-photon systems; however, the time-reversal method is supposed to be applicable to all kinds of entangled two-photon systems in theory.
Here, for the first time to the best of our knowledge, we apply the time-reversal method to position--wavevector-entangled two-photon systems.
We focus on the two-photon Young interferometer \cite{PhysRevLett.82.2868,PhysRevLett.85.2733,PhysRevLett.87.013602,kawabe2007quantum,PhysRevLett.112.223602} and the two-photon beam focusing system \cite{xu2015experimental}, and we experimentally demonstrate that the time-reversed versions of these two systems classically reconstruct the same interference patterns as those in the time-forward systems.

This paper is organized as follows. 
In Sec.~\ref{sec:time-reversal-method}, we present the theory of the time-reversal method, particularly for two-photon systems.
In Sec.~\ref{sec:time-reversal-method-1} and \ref{sec:time-reversal-method-2}, we describe our two experiments.
First, we demonstrate the time-reversal method for the two-photon Young interferometer in Sec.~\ref{sec:time-reversal-method-1}.
Subsequently, we demonstrate the time-reversal method for the two-photon beam focusing system in Sec.~\ref{sec:time-reversal-method-2}.
Finally, we summarize the findings of our study in Sec.~\ref{sec:conclusion}.


\section{Time-reversal method}\label{sec:time-reversal-method}

In this section, we present the theory underlying the time-reversal method for two-photon systems.
We begin by reviewing the time-reversal symmetry of quantum mechanics.
Let us consider the process where initial state $\ket{\ii}$ evolves with unitary operator $\hat{U}$ and is projected onto final state $\ket{\ff}$ by a measurement apparatus.
The projection probability is given by $|\bracketii{\ff}{\hat{U}}{\ii}|^2$.
In its time-reversed process, where initial state $\ket{\ff}$ evolves with $\hat{U}^{-1}$ and is projected onto final state $\ket{\ii}$, the projection probability is given by $|\bracketii{\ii}{\hat{U}^{-1}}{\ff}|^2$.
This probability is equal to that of the time-forward process due to the conjugate transposition $\bracketii{\ff}{\hat{U}}{\ii}=\bracketii{\ii}{\hat{U}^{\dag}}{\ff}^*$ and unitarity $\hat{U}^\dag=\hat{U}^{-1}$.

\begin{figure}[t]
\begin{center}
\includegraphics[width=8.5cm]{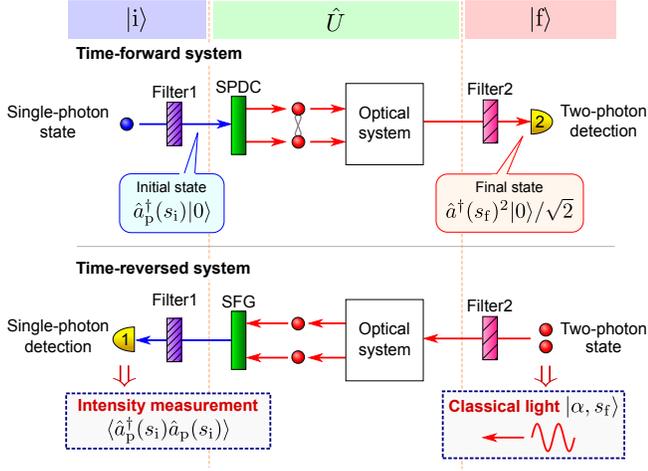}
\caption{Schematics of two-photon system with two-photon detection in a single output mode (upper panel) and its time-reversed system (lower panel).
Filter1 and filter2 function as pre-selection of $\ket{\ii}$ and post-selection of $\ket{\ff}$ in the time-forward system, respectively. 
In the time-reversed system, the two-photon detection and the single-photon input state in the time-forward system are replaced by the two-photon input state and the single-photon detection, respectively; here, filter1 and filter2 function as post-selection of $\ket{\ii}$ and pre-selection of $\ket{\ff}$, respectively. 
The time-reversed system can be prepared by the use of a classical optical system in which classical input light and intensity measurement are employed.
}\label{fig:1}
\end{center}
\end{figure}

We next consider the time-reversal symmetry of two-photon systems including SPDC.
In particular, here, we here treat two-photon systems in which the initial state is a pure pump-photon state and the final state is a pure two-photon state in a single-output mode.
The two-photon Young interferometer and the two-photon beam focusing system, which we address later in this paper, fall under this case. 
Figure~\ref{fig:1} shows the schematics of such a two-photon system (upper panel) and its time-reversed system (lower panel). 
In the time-forward system, the initial state $\ket{\ii}=\hat{a}\sub{p}^\dag(s\sub{i})\ket{0}$ is pre-selected by filter1, where $\hat{a}\sub{p}^\dag(s)$ denotes the creation operator of a pump photon parametrized by parameter $s$, which can denote position, wavenumber, time, or frequency.
We assume that $\hat{a}\sub{p}^\dag(s)$ satisfies the following commutation relation: $[\hat{a}\sub{p}(s),\hat{a}\sub{p}^\dag(s')]=\delta(s-s')$. 
The input photon is converted into a photon pair by SPDC, and the photon pair passes through the optical system.
The overall time evolution is denoted by unitary operator $\hat{U}$, with which creation operator $\hat{a}\sub{p}^\dag(s\sub{i})$ evolves into
\begin{align}
\hat{U}\hat{a}\sub{p}^\dag(s\sub{i})\hat{U}^\dag=\int\dd s\int\dd s' f(s,s';s\sub{i})\hat{a}^\dag(s)\hat{a}^\dag(s'),
\end{align}
where $\hat{a}^\dag(s)$ denotes the creation operator of a down-converted photon parametrized by $s$.
Coefficient $f(s,s';s\sub{i})$ satisfies the unitary condition $\hat{U}^\dag\hat{U}=\hat{1}$; we can assume that $f(s, s'; s\sub{i}) = f(s', s; s\sub{i})$ because of the exchange symmetry of bosons.
If $f(s,s';s\sub{i})$ cannot be factorized in the form of $g(s;s\sub{i})g(s';s\sub{i})$, the two photons are entangled.
The photon pair is finally post-selected by filter2 and detected by means of two-photon detection in a single-output mode.
This detection is interpreted as the projection onto final state $\ket{\ff}=\hat{a}^\dag(s\sub{f})^2\ket{0}/\sqrt{2}$. 
The detection probability is given by
\begin{align}
|\bracketii{\ff}{\hat{U}}{\ii}|^2&=\frac{1}{2}|\bracketii{0}{\hat{a}(s\sub{f})^2\hat{U}\hat{a}\sub{p}^\dag(s\sub{i})\hat{U}^\dag}{0}|^2\no\\
&=2|f(s\sub{f},s\sub{f};s\sub{i})|^2.\label{eq:9}
\end{align}

In the time-reversed system, the initial state is a two-photon state $\ket{\ff}$ in the input mode corresponding to the output mode in the time-forward system.
The pair of photons experience the optical system in reverse, and the pair is converted into a single photon by sum-frequency generation (SFG); the overall time evolution is $\hat{U}^{-1}$.
The up-converted photon is finally projected onto final state $\ket{\ii}$.
The detection probability $|\bracketii{\ii}{\hat{U}^{-1}}{\ff}|^2$ is equal to that of the time-forward system expressed in Eq.~(\ref{eq:9}) because of the time-reversal symmetry.

Interestingly, the time-reversed system of a two-photon system can be prepared by the use of a classical optical system.
To examine this possibility, here, we consider a classical time-reversed system in which the input two-photon state $\ket{\ff}$ is replaced by a coherent state $\ket{\alpha,s\sub{f}}$ defined as 
\begin{align}
\ket{\alpha,s\sub{f}}:=\ee^{-|\alpha|^2/2}\sum_{n=0}^{\infty}\frac{[\alpha\hat{a}^\dag(s\sub{f})]^n}{n!}\ket{0},
\end{align}
and the single-photon detection is replaced by an optical intensity measurement.
The result of the optical intensity measurement is represented as an expectation value of the photon number:
\begin{align}
\bracket{\hat{a}\sub{p}^\dag(s\sub{i})\hat{a}\sub{p}(s\sub{i})}
&=\bracketii{\alpha,s\sub{f}}{\hat{U}\hat{a}\sub{p}^\dag(s\sub{i})\hat{a}\sub{p}(s\sub{i})\hat{U}^\dag}{\alpha,s\sub{f}}\no\\
&=\left\|\hat{U}\hat{a}\sub{p}(s\sub{i})\hat{U}^\dag\ket{\alpha,s\sub{f}}\right\|^2\no\\
&=\left\|\int\dd s\int\dd s' f^*(s,s';s\sub{i})\hat{a}(s)\hat{a}(s')\ket{\alpha,s\sub{f}}\right\|^2\no\\
&=\left\|\alpha^2 f^*(s\sub{f},s\sub{f};s\sub{i})\ket{\alpha,s\sub{f}}\right\|^2\no\\
&\propto|f(s\sub{f},s\sub{f};s\sub{i})|^2,
\end{align}
where we use the relation $\hat{a}(s)\ket{\alpha,s\sub{f}}=\alpha\delta(s-s\sub{f})\ket{\alpha,s\sub{f}}$.
As can be observed above, the measured intensity distribution in the classical time-reversed system exhibits the same pattern as the two-photon detection pattern in the time-forward system. 

We remark on the two following points regarding the time-reversal method for two-photon systems.
First, we treat two-photon systems that allow two-photon detection in a single-output mode, such as the two-photon Young interferometer and the two-photon beam focusing system.
More general two-photon systems have multiple output modes, and each of the two photons is detected in a different output mode. 
In this case, time-reversed classical systems need a conditional SFG, which eliminates the two photons not corresponding to the entangled photon pairs in the time-forward system.
The details of the time-reversal method for general two-photon systems are provided in Appendix~\ref{sec:time-reversed-method}.
Second, although the time-reversed classical system can reconstruct the two-photon detection patterns in the time-forward system, this fact does not mean that the quantum-optical phenomena are realized classically.
The time-reversed system is completely classical and merely exhibits the same intensity pattern as the two-photon detection patterns.

\section{Classical reconstruction of two-photon Young interference}\label{sec:time-reversal-method-1}

In this section, we apply the time-reversal method to the two-photon Young interferometer \cite{PhysRevLett.82.2868,PhysRevLett.85.2733,PhysRevLett.87.013602,kawabe2007quantum,PhysRevLett.112.223602}.
The two-photon Young interference fringe has half the period of classical Young interference fringes, which are caused by position--wavevector entanglement of photon pairs.
In Sec.~\ref{sec:time-forward-time}, we introduce the time-forward two-photon Young interferometer, and in Sec.~\ref{sec:time-reversed-system}, we construct its time-reversed system in accordance with the time-reversal method.
In Sec.~\ref{sec:experiment-results}, we experimentally demonstrate that the two-photon Young interference pattern can be reconstructed classically in the time-reversed system.  

\subsection{Time-forward system}\label{sec:time-forward-time}

First, we introduce the time-forward two-photon Young interferometer.
Hereafter, we assume that creation operators parametrized by lateral position $x$ and lateral wavenumber $k_x$ are, respectively, described by small letter $\hat{a}^\dag(x)$ and capital letter $\hat{A}^\dag(k_x)$. 
Two-photon Young interference is typically realized as follows. 
We prepare the two-photon NOON state 
$[\hat{a}^\dag(x_1)^2+\hat{a}^\dag(-x_1)^2]\ket{0}$ at a double slit, where
$\pm x_1$ denote the lateral positions of the left and right slits, respectively; here, we ignore the normalization constant.
Next, we focus two photons with incidence angles $\pm\theta$, respectively, and we detect them with a two-photon detector positioned in the focal plane.
Upon moving the position of the two-photon detector, the two-photon counting rate yields the interference fringe with period $\lambda/(4\sin\theta)$, where $\lambda$ denotes the wavelength of the incidence photons.
On the other hand, the classical Young interference with the same incidence angles $\pm\theta$ exhibits period $\lambda/(2\sin\theta)$; therefore, the period of the two-photon Young interference fringe is half of that of the classical one and can overcome the diffraction limit.

\begin{figure}[t]
\begin{center}
\includegraphics[width=8.5cm]{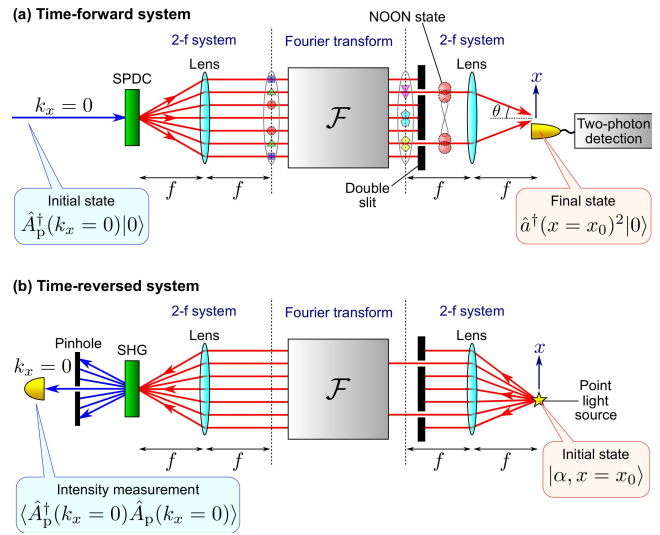}
\caption{Schematics of (a) time-forward and (b) time-reversed two-photon Young interferometers.
}\label{fig:2}
\end{center}
\end{figure}

Two-photon Young interference has been observed in several experimental configurations \cite{PhysRevLett.82.2868,PhysRevLett.85.2733,PhysRevLett.87.013602,kawabe2007quantum,PhysRevLett.112.223602}, and we next consider the setup shown in Fig.~\ref{fig:2}(a). 
Here, we consider only the one-dimensional lateral distribution of the wavefunctions.
In this setup, initial state $\ket{\ii}$ is a single pump photon state (wavelength $\lambda/2$) with a single lateral wavenumber $(k_x=0)$: $\ket{\ii}=\hat{A}\sub{p}^\dag(0)\ket{0}$, where $\hat{A}\sub{p}^\dag(k_x)$ denotes the creation operator of a pump photon with lateral wavenumber $k_x$.
This state is first down-converted into the position--wavevector-entangled two-photon state $\int\dd k_x\hat{A}^\dag(k_x)\hat{A}^\dag(-k_x)\ket{0}=\int\dd x\hat{a}^\dag(x)^2\ket{0}$ by SPDC, in which the lateral positions of the two photons are positively correlated.
Next, the photon pair is subjected to the first 2-$f$ system and an optical Fourier transform such as long-distance free-space propagation.
In total, the photon pair is subjected to a magnifying optical system and the same two-photon state $\int\dd x\hat{a}^\dag(x)^2\ket{0}$ appears in front of the double slit 
(The reason we employ this setup is to tightly focus the fundamental light into the nonlinear crystal and to generate high-power sum-frequency light in the time-reversed system [Fig.~\ref{fig:2}(b)] as mentioned later).
The photon pair after passing through the double slit is represented by a two-photon NOON state: $[\hat{a}^\dag(x_1)^2+\hat{a}^\dag(-x_1)^2]\ket{0}$ \cite{footnote2}.
Finally, the NOON state is Fourier-transformed by the second 2-$f$ system.
Creation operator $\hat{a}^\dag(x)$ is transformed into 
\begin{align}
\mathcal{F}[\hat{a}^\dag(x')]\left(\frac{2\pi x}{f\lambda}\right)
=\int\dd x'\hat{a}^\dag(x')\exp\left(\frac{-\ii 2\pi xx'}{f\lambda}\right),
\end{align}
where $f$ represents the focal length of the second lens and $\mathcal{F}[g(x')](k):=\int\dd x'g(x')\ee^{-\ii x'k}$ denotes the Fourier transform.
In total, initial state $\ket{\ii}$ is converted into
\begin{align}
\ket{\ii'}
:=\int\dd x'\int\dd x''\cos\left[\frac{2\pi x_1(x'+x'')}{f\lambda}\right]\hat{a}^\dag(x')\hat{a}^\dag(x'')\ket{0}.
\end{align}
The photon pair is finally detected at lateral position $x=x_0$ on the focal plane.
Final state $\ket{\ff}$ is represented as $\ket{\ff}=\hat{a}^\dag(x_0)^2\ket{0}$, and therefore, the two-photon counting probability $P(x_0)$ is given by
\begin{align}
P(x_0)\propto|\bracketi{\ff}{\ii'}|^2\propto\frac{1}{2}\left[1+\cos\left(\frac{8\pi x_1x_0}{f\lambda}\right)\right].\label{eq:11}
\end{align}
This probability distribution $P(x_0)$ has period $\lambda/(4\tan\theta)$, where $\tan\theta=x_1/f$.
When $\theta\ll 1$, $\tan\theta\simeq\sin\theta$, and therefore, this period agrees with the value $\lambda/(4\sin\theta)$ mentioned in the previous paragraph.
In this manner, the NOON states are prepared, and two-photon Young interference is subsequently observed.

\subsection{Time-reversed system}\label{sec:time-reversed-system}

We next consider the time-reversed two-photon Young interferometer.
In accordance with the time-reversal method, the classical time-reversed system is constructed as shown in Fig.~\ref{fig:2}(b).
In the time-reversed system, the movable two-photon detector in the time-forward system is replaced by the movable classical point light source $\ket{\alpha,x_0}$.
Further, the preparation of the pump photons with a single lateral wavenumber in the time-forward system is replaced by a pinhole at $x=0$ as a lateral wavevector filter and optical intensity measurement, which is represented as $\bracket{\hat{A}\sub{p}^\dag(0)\hat{A}\sub{p}(0)}$.
Upon shifting lateral position $x_0$ of the point light source, the measured intensity distribution exhibits the same interference pattern as the time-forward system Eq.~(\ref{eq:11}) because of the time-reversal symmetry.
The detailed calculation of this interference pattern is provided in Appendix~\ref{sec:young}.

\subsection{Experiments and results}\label{sec:experiment-results}

\begin{figure}[t]
\begin{center}
\includegraphics[width=8.5cm]{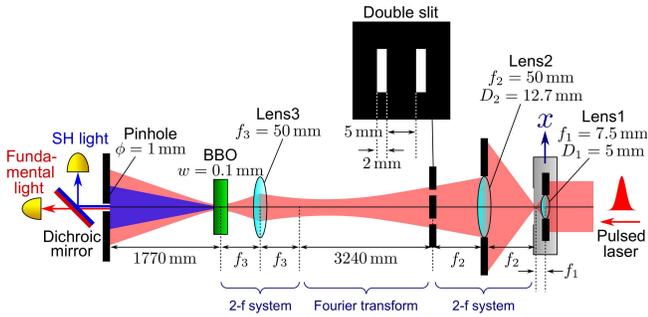}
\caption{Experimental setup of time-reversed two-photon Young interferometer. 
}\label{fig:3}
\end{center}
\end{figure}

We experimentally demonstrate that the time-reversed two-photon Young interferometer reconstructs the same interference patterns as the time-forward one.
The experimental setup that implements the time-reversed system is shown in Fig.~\ref{fig:3}.
In the study, we used a pulsed laser (Menlo Systems, C-Fiber 780; central wavelength 780\,nm, pulse width 120\,fs, average power 120\,mW, repetition rate 100\,MHz) as a light source.
The beam was collimated with a beam width of 7.4\,mm and was focused by lens1 (focal length $f_1=7.5$\,mm, diameter $D_1=5$\,mm) to prepare a pseudo point light source (spot size 1.49\,$\upmu$m, depth-of-focus 4.46\,$\upmu$m).
The lateral position $x$ of the point light source can be changed by moving the stage mounting lens1.
The second 2-$f$ system was implemented by lens2 (focal length $f_2=50$\,mm, diameter $D_2=$12.7\,mm).
Beyond the 2-$f$ system, the double slit shown in Fig.~\ref{fig:3} was inserted.
The subsequent Fourier transform was realized via long-distance (3240\,mm) free-space propagation.
The other 2-$f$ system was implemented by means of lens3 (focal length $f_3=50$\,mm).
The beam was focused into a 0.1-mm-length $\upbeta$-barium borate (BBO) crystal for type-I SHG. 
The fundamental and sum-frequency (SH) light passed through a pinhole (diameter 1\,mm), which allows transmission of a narrow lateral wavevector component, and were divided by a dichroic mirror.
The optical powers of the fundamental and SH light were measured by a Si photodetector (Thorlabs, PDA100A) and a Si femtowatt detector (Thorlabs, PDF10A/M), respectively.
We measured the optical powers of the fundamental and SH light at various lateral positions $x$ of the point light source.

\begin{figure}[t]
\begin{center}
\includegraphics[width=8.5cm]{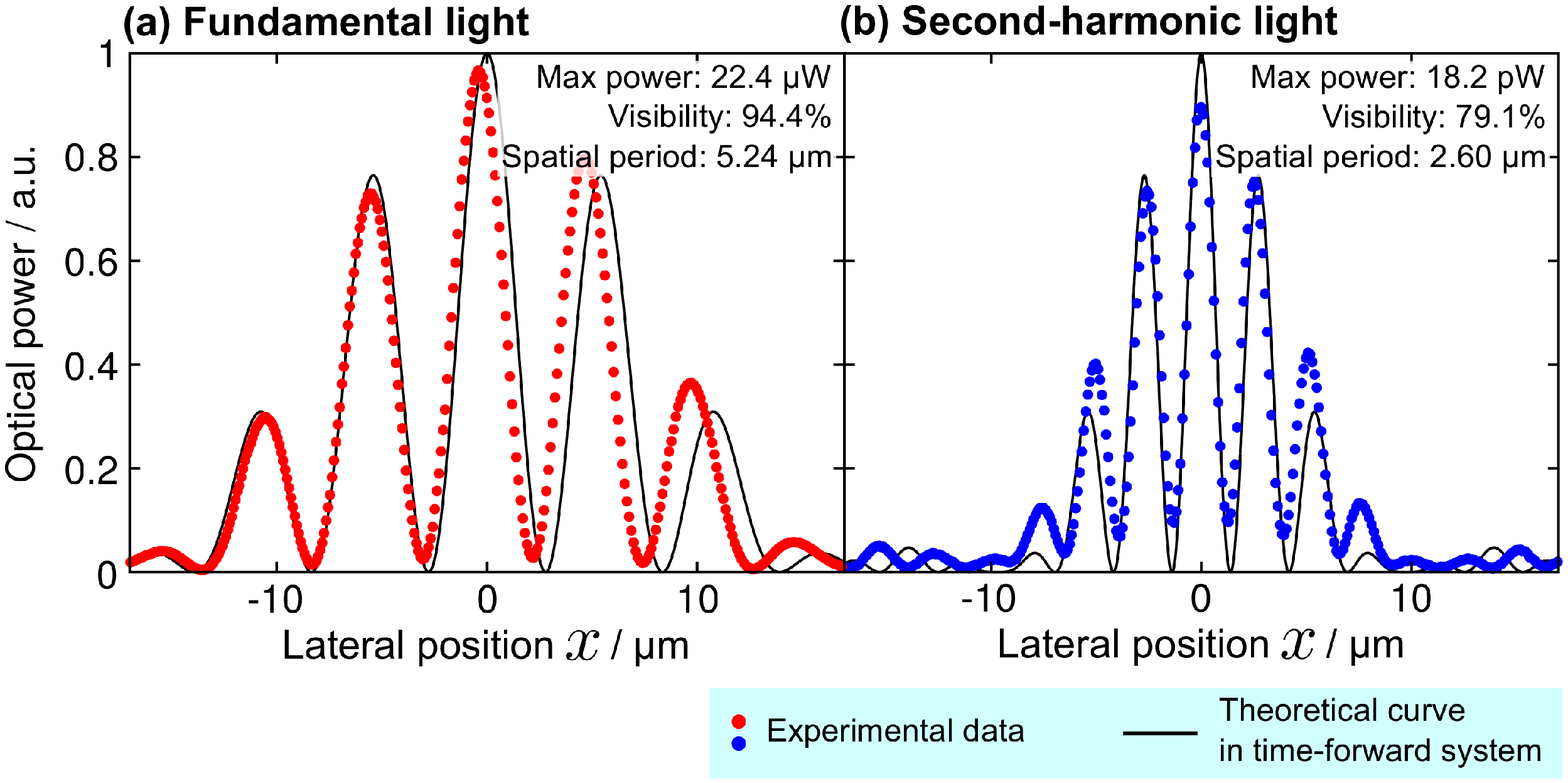}
\caption{Experimental results of the time-reversed two-photon Young interferometer for fundamental light (a) and second-harmonic light (b).
The optical powers in the longitudinal axes are normalized by the measured maximum powers of 22.4\,$\upmu$W (a) and 18.2\,pW (b), respectively.
The solid curves represent the theoretical curves corresponding to the time-forward classical and two-photon Young interferometers with the same experimental conditions as this set of experiments.
}\label{fig:4}
\end{center}
\end{figure}

The experimental results for the fundamental and SH light are shown in Figs.~\ref{fig:4}(a) and (b), respectively.
The solid curves denote the theoretical curves in the time-forward classical and two-photon Young interferometers under the same experimental conditions as this set of experiments (double slit and incidence angles).
Both results show good agreement with the theoretical curves of the time-forward systems, and in particular, the result for the SH light reconstructs the same interference pattern as the time-forward system.
Therefore, the time-reversal method for the two-photon Young interferometer is demonstrated.
We note that the observed decline in the visibility was caused by the difference between the power transmitted passing through the left and light slits.


\section{Classical reconstruction of two-photon focused beam spot}\label{sec:time-reversal-method-2}

In this section, we describe our application of the time-reversal method to the two-photon beam focusing system \cite{xu2015experimental}.
The two-photon focused beam spot has sub-diffraction-limited spot size and depth-of-focus, which are caused by position--wavevector entanglement of photon pairs.
In Sec.~\ref{sec:time-forward-system}, we introduce the time-forward two-photon beam focusing system, and in Sec.~\ref{sec:time-reversed-system-1} we construct its time-reversed system in accordance with the time-reversal method.
In Sec.~\ref{sec:experiment-results-1}, we experimentally demonstrate that the two-photon focused beam spot can be reconstructed classically in the time-reversed system.

\subsection{Time-forward system}\label{sec:time-forward-system}

\begin{figure}[t]
\begin{center}
\includegraphics[width=8.5cm]{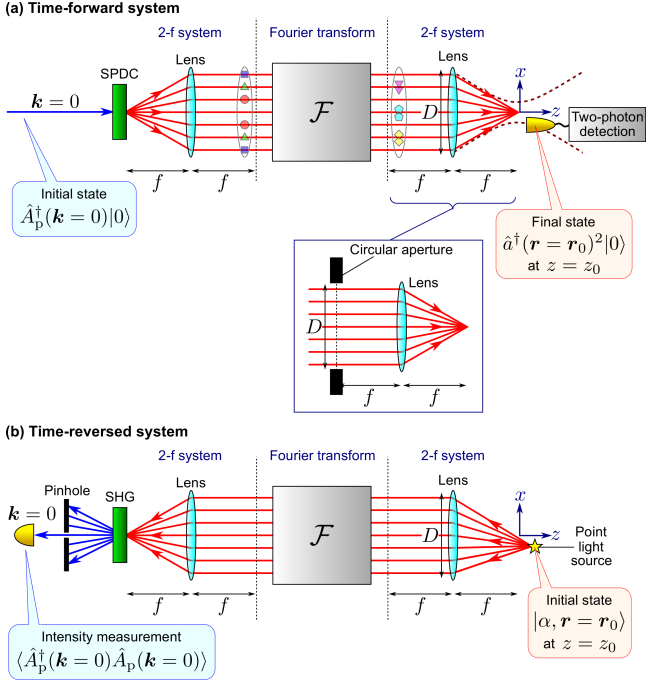}
\caption{Schematics of (a) time-forward and (b) time-reversed two-photon beam focusing system. 
(Inset) Approximated 2-$f$ system constructed by lens with finite diameter $D$.
}\label{fig:5}
\end{center}
\end{figure}

Here, we introduce the time-forward two-photon beam focusing system shown in Fig.~\ref{fig:5}(a).
We consider the two-dimensional lateral distribution of the wavefunctions unlike in Sec.~\ref{sec:time-forward-time}.
The two-photon focusing beam spot is realized by focusing the position--wavevector-entangled photon pairs $\int\dd\vct{r}\hat{a}^\dag(\vct{r})^2\ket{0}$, where $\vct{r}=(x,y)$ denotes the lateral position vector.
This state is positively correlated along the lateral position.
This entangled two-photon state can be prepared by using the same optical system as that used for the two-photon Young interferometer described in Sec.~\ref{sec:time-reversal-method-1}.
The difference between the system shown in Fig.~\ref{fig:5}(a) and that in Fig.~\ref{fig:2}(a) is that the double slit is removed and the two-photon detector can move along the lateral ($x$) and axial ($z$) directions.
We now consider that the lens in the second (right) 2-$f$ system has a finite diameter $D$. 
If the beam is collimated sufficiently at the back focal plane of the 2-$f$ system, this 2-$f$ system can be approximated by a 2-$f$ system constructed by a lens with an infinite diameter and a circular aperture with diameter $D$ at the back focal plane, as shown in the inset of Fig.~\ref{fig:5}. 
In the following discussion, we assume this approximation for simplicity.

Next, we derive the distribution of the two-photon focused beam spot.
Initial state $\ket{\ii}$ is the same as that of the two-photon Young interferometer described in Sec.~\ref{sec:time-forward-time}: $\ket{\ii}=\hat{A}\sub{p}^\dag(0)\ket{0}$, where the parameter of $\hat{A}\sub{p}^\dag(\vct{k})$ is the lateral wavevector $\vct{k}=(k_x,k_y)$.
This state is subjected to the optical system shown in Fig.~\ref{fig:5}(a), and subsequently, the two-photon state at the back focal plane of the second (right) 2-$f$ system can be represented as
\begin{align}
\int_{|\vct{r}|\leq\frac{D}{2}}\dd\vct{r}
\hat{a}^\dag(\vct{r})^2\ket{0}.\label{eq:6}
\end{align}
After $(2f+z)$-distance propagation through the 2-$f$ system ($z$ denoting the distance from the focal plane), $\hat{a}^\dag(\vct{r})$ is transformed into
\begin{align}
(f+z)\exp\left(\frac{-\ii\pi z|\vct{r}|^2}{f^2\lambda}\right)
\int\dd\vct{r}'
\exp\left(\frac{-\ii 2\pi\vct{r}\cdot\vct{r}'}{f\lambda}\right)
\hat{a}^\dag(\vct{r}')
\end{align}
as per the Fresnel approximation.
In total, the initial state $\ket{\ii}$ is transformed into
\begin{align}
\ket{\ii''}:=&
(f+z)^2
\int_{|\vct{r}|\leq\frac{D}{2}}\dd\vct{r}
\exp\left(\frac{-\ii 2\pi z|\vct{r}|^2}{f^2\lambda}\right)\no\\
&\times
\left[
\int\dd\vct{r}'
\exp\left(\frac{-\ii 2\pi\vct{r}\cdot\vct{r}'}{f\lambda}\right)
\hat{a}^\dag(\vct{r}')
\right]^2
\ket{0}.
\end{align}
The photon pair is finally detected at lateral position $\vct{r}=\vct{r}_0$ and an axial position from the focal plane $z=z_0$. 
Final state $\ket{\ff}$ is represented as $\hat{a}^\dag(\vct{r}_0)^2\ket{0}$ at $z=z_0$, and therefore, the two-photon counting probability $P(\vct{r}_0,z_0)$ is given by
\begin{align}
&P(\vct{r}_0,z_0)\propto|\bracketi{\ff}{\ii''}|^2\no\\
&\propto
\Bigg|(f+z_0)^2
\int_{|\vct{r}|\leq\frac{D}{2}}\dd\vct{r}
\exp
\left(\frac{-\ii 2\pi z_0|\vct{r}|^2}{f^2\lambda}\right)\no\\
&\hspace{3cm}
\times\exp
\left(\frac{-\ii 4\pi\vct{r}\cdot\vct{r}_0}{f\lambda}\right)
\Bigg|^2.\label{eq:7}
\end{align}

When $z_0=0$, the two-photon counting probability is
\begin{align}
P(\vct{r}_0,0)\propto\mathrm{somb}^2
\left(\frac{2\pi D|\vct{r}_0|}{f\lambda}\right),
\end{align}
where $\mathrm{somb}(x):=J_1(x)/x$ represents the sombrero function and $J_n(x)$ denotes the $n$-th order Bessel function of the first kind.
The full-width at half-maximum (FWHM) of this distribution in the lateral direction is given by $1.62\lambda f/(\pi D)$.
On the other hand, the classical beam spot focused by the same focusing lens has the intensity distribution of $\mathrm{somb}^2[\pi D|\vct{r}_0|/(f\lambda)]$, which corresponds to a the spot size of $3.23\lambda f/(\pi D)$ FWHM. 
Therefore, the width of the two-photon focusing beam spot in the lateral direction is half the classical one and overcomes the diffraction limit.

When $\vct{r}_0=0$, the two-photon counting probability is
\begin{align}
P(0,z_0)\propto&
(f+z_0)^4\mathrm{sinc}^2
\left(\frac{\pi D^2z_0}{4f^2\lambda}\right),
\end{align}
where $\mathrm{sinc}(x):=\sin(x)/x$.
Assuming that $z_0\ll f$ and $(f+z_0)^4$ can be zero-order-approximated by $(f+z_0)^4\sim f^4$, the FWHM of this distribution in the axial direction is given by $11.1\lambda f^2/(\pi D^2)$.
The classical beam spot focused by the same focusing lens has the intensity distribution of $(f+z_0)^4\mathrm{sinc}^2[\pi D^2z_0/(8f^2\lambda)]$, which corresponds to a the depth-of-focus of $3.23\lambda f/(\pi D)$ FWHM with the same approximation. 
Therefore, the width of the two-photon focusing beam spot in the axial direction is also half the classical one and overcomes the diffraction limit.

\subsection{Time-reversed system}\label{sec:time-reversed-system-1}

We next consider the time-reversed two-photon beam focusing system shown in Fig.~\ref{fig:5}(b).
As is the case in the two-photon Young interferometer, the movable two-photon detector and the preparation of the pump photons with a single lateral wavenumber in the time-forward system are replaced by the movable classical point light source $\ket{\alpha,\vct{r}_0}$ at $z=z_0$ and a pinhole followed by optical intensity measurement, which is represented as $\bracket{\hat{A}\sub{p}^\dag(0)\hat{A}\sub{p}(0)}$, respectively.
The point light source can be moved along the lateral ($x$) and axial ($z$) directions.
Due to time-reversal symmetry, the measured intensity distribution for various $x$ and $z$ values shows the same pattern as the time-forward two-photon beam focusing system corresponding to Eq.~(\ref{eq:7}).
The detailed calculation of this intensity distribution is provided in Appendix~\ref{sec:spot}. 

\subsection{Experiments and results}\label{sec:experiment-results-1}

\begin{figure}[t]
\begin{center}
\includegraphics[width=8.5cm]{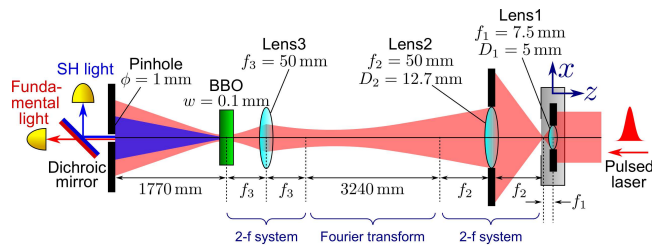}
 \caption{Experimental setup of the time-reversed two-photon beam focusing system. 
}\label{fig:6}
\end{center}
\end{figure}

We experimentally demonstrate that the time-reversed two-photon beam focusing system reconstructs the same optical power distributions as the two-photon detection patterns in the time-forward system.
The experimental setup is shown in Fig.~\ref{fig:6}.
This setup is identical to that shown in Fig.~\ref{fig:3} except that the double slit is removed and the stage mounting lens1 can move along the lateral $(x)$ and axial $(z)$ directions.  
In this study, we measured the optical power of the fundamental and SH light beams at various lateral and longitudinal positions $x$ and $z$ of the point light source.

\begin{figure*}[t]
\begin{center}
\includegraphics[width=15cm]{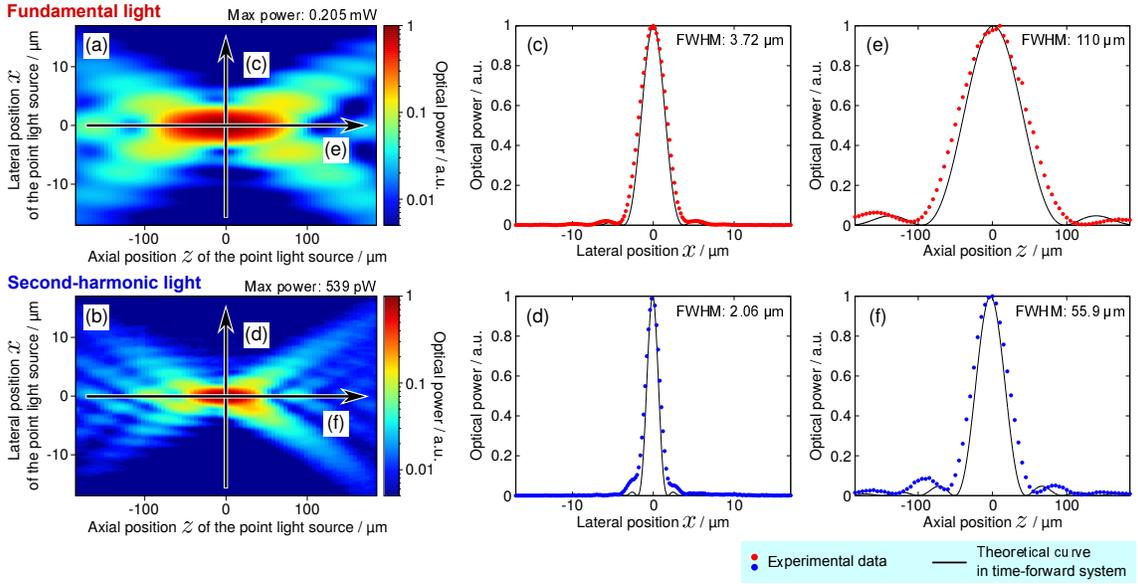}
\caption{
Experimental results of the time-reversed two-photon beam focusing system for the fundamental light (upper panels) and the second-harmonic light (lower panels). 
(a), (b) Optical power distributions at lateral and axial positions $x$ and $z$ of the point light source. 
The optical power in the color bars is normalized by the measured maximum power of 0.205\,mW (a) and 539\,pW (b), respectively.
(c), (d) Optical power distributions at $x$ when $z=0$. 
(e), (f) Optical power distributions at $z$ when $x=0$.
The solid curves denote the theoretical curves corresponding to the time-forward classical and two-photon beam focusing systems with the same experimental conditions as this set of experiments.
}\label{fig:7}
\end{center}
\end{figure*}

The experimental results are shown in Figs.~\ref{fig:7}(a) to (f).
The upper and lower panels illustrate the results for the fundamental and SH light, respectively.
Figures~\ref{fig:7}(a) and (b) depict the measured optical power distributions at various lateral and axial positions $x$ and $z$ of the point light source. 
Each of the distributions is  similar to that of the focused beam spot in the time-forward system, and the distribution of the SH light is smaller than that of the fundamental light. 
Figures~\ref{fig:7}(c) and (d) show the optical power distributions at lateral position $x$ when $z=0$, and Figs.~\ref{fig:7}(e) and (f) show those at axial position $z$ when $x=0$. 
The solid curves represent the theoretical curves of the time-forward classical and two-photon beam focusing systems with the same experimental conditions (diameter and focal length of lens2) as this set of experiments.
Each of the results shows good agreement with the theoretical curves of the time-forward systems, and in particular, the result for the SH light reconstructs the same optical power distributions as the two-photon detection patterns in the time-forward system.
In Figs.~\ref{fig:7}(a) and (b), the distributions are not symmetric about the longitudinal axis at $z=0$\,$\upmu$m; this is because lens2's numerical aperture (NA) is large, and therefore, this system deviates from the paraxial approximation condition. 
The experimental results exhibit a slightly wider spread than the theoretical curves due to the aberration effect of lens2.
The time-reversal method for two-photon beam focusing system is thus demonstrated.






\section{Conclusion}\label{sec:conclusion}

We experimentally demonstrated the time-reversal method for the two-photon Young interferometer and the two-photon beam focusing system.
These time-reversed systems classically reconstructed the same interference patterns as those of the time-forward systems.
To the best of our knowledge, our study is the first to demonstrate the time-reversal method for quantum-optical phenomena caused by two-photon entanglement except for time--frequency entanglement; our study particularly addresses quantum-optical phenomena arising due to position--wavevector entanglement.
The theory and experiments presented in this study can form the basis for applying the time-reversal method to a wider range of quantum-optical phenomena.
It is expected that the time-reversal method can provide the approach to classically realize application techniques based on quantum-optical phenomena, such as sub-Rayleigh imaging \cite{PhysRevA.79.013827,PhysRevLett.105.163602,xu2015experimental}.

\section*{Acknowledgements}

This research is supported by JSPS KAKENHI Grant Number 16K17524.

\appendix


\section{Time-reversed method for general two-photon systems}\label{sec:time-reversed-method}

\begin{figure}[t]
\begin{center}
\includegraphics[width=8.5cm]{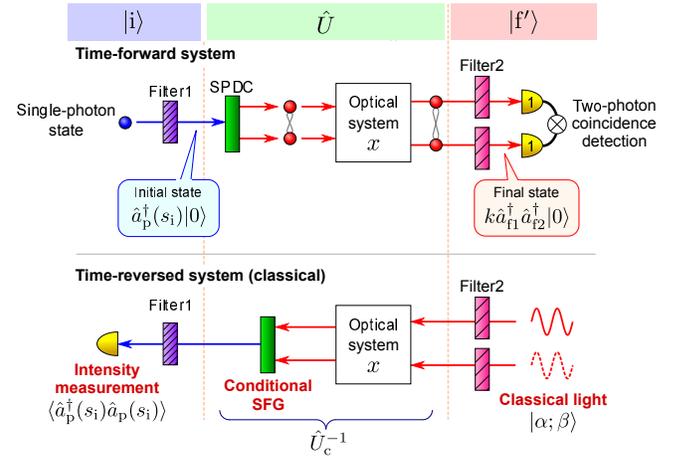}
\caption{Schematics of general two-photon system in which each of the two photons is detected in different output modes (upper panel) and its classical time-reversed system (lower panel).
The time-reversed system can be prepared by the use of a classical optical system in which classical input light, conditional SFG, and intensity measurement are employed.
}\label{fig:8}
\end{center}
\end{figure}

In Sec.~\ref{sec:time-reversal-method}, we introduced the time-reversal method for two-photon systems with two-photon detection in a single-output mode. 
Here, we present the theory underlying the time-reversal method for general two-photon systems, i.e., two-photon systems in which the two output photons are generally detected in different output modes.
Figure~\ref{fig:8} shows the schematics of such a two-photon system (upper panel) and its classical time-reversed system (lower panel).
For example, the HOM interferometer \cite{PhysRevLett.59.2044} has two output ports, and each of the two photons is detected at each of the output ports.
Another example is the modified HOM interferometer that exhibits an HOM peak \cite{PhysRevLett.76.4656}.
This interferometer has only one output port, but two photons with various time differences are detected at this port; therefore, it can be considered that each of the two photons is detected in different time modes.
Such a final pure state is generally described as $\ket{\ff'}=k\hat{a}\sub{f1}^\dag\hat{a}\sub{f2}^\dag\ket{0}$, where $\hat{a}_{\mathrm{f}i}^\dag:=\int\dd s\psi_{\mathrm{f}i}(s)\hat{a}^\dag(s)$ [$\psi_{\mathrm{f}i}(s)$ denotes a normalized wavefunction] for $i=1,2$, and $k$ is the normalization factor given as $k:=[1+|\int\dd s\psi\sub{f1}^*(s)\psi\sub{f2}(s)|^2]^{-1/2}$.
When $\psi\sub{f1}(s)=\psi\sub{f2}(s)$, $\ket{\ff'}$ denotes the two-photon state in a single mode, as considered in Sec.~\ref{sec:time-reversal-method}.
The detection probability is given by
\begin{align}
&|\bracketii{\ff'}{\hat{U}}{\ii}|^2=4k^2\left|\int\dd s_1\psi^*\sub{f1}(s_1)\int\dd s_2\psi^*\sub{f2}(s_2)f(s_1,s_2;s\sub{i})\right|^2.\label{eq:10}
\end{align}

The classical time-reversed system of the general two-photon system can be constructed in the following manner.
First, we prepare the input coherent state $\ket{\alpha;\beta}$, which corresponds to $\ket{\ff'}$, defined as 
\begin{align}
\ket{\alpha;\beta}:=\ee^{-(|\alpha|^2+|\beta|^2)/2}\sum_{m,n=0}^{\infty}\frac{(\alpha\hat{a}\sub{f1}'^\dag)^m(\beta\hat{a}\sub{f2}'^\dag)^n}{m!n!}\ket{0},
\end{align}
where $\hat{a}_{\mathrm{f}i}'^\dag:=\int\dd s\psi_{\mathrm{f}i}(s)\hat{a}_i^\dag(s)$, for $i=1,2$, and $\hat{a}_i^\dag(s)$ satisfies the following commutation relation: $[\hat{a}_i(s),\hat{a}_j^\dag(s')]=\delta(s-s')\delta_{ij}$.
The subscripts of creation operators $\hat{a}_1^\dag$ and $\hat{a}_2^\dag$ are introduced in order to distinguish the two photons from the different input modes.
Next, the SFG is designed such that the up-converted light includes only the contributions of $\hat{a}_1^\dag(s)\hat{a}_2^\dag(s)$; in other words, the contributions of two photons from the same input mode $\hat{a}_i^\dag(s)^2$ are eliminated.
For the case of polarization, for instance, such a conditional SFG can be realized by type-II SFG.
Due to the conditional SFG, the overall time evolution of this system is represented by $\hat{U}\sub{c}^{-1}=\hat{U}\sub{c}^\dag$, with which the creation operator $\hat{a}\sub{p}^\dag(s\sub{i})$ is transformed into
\begin{align}
\hat{U}\sub{c}\hat{a}\sub{p}^\dag(s\sub{i})\hat{U}\sub{c}^\dag
=\int\dd s_1\int\dd s_2f(s_1,s_2;s\sub{i})\hat{a}_1^\dag(s_1)\hat{a}_2^\dag(s_2).
\end{align}
In previous studies, such a conditional SFG has been realized by utilizing the degrees of freedom of frequency \cite{kaltenbaek2008quantum,Lavoie2009,mazurek2013dispersion,PhysRevLett.102.243601} and polarization \cite{PhysRevA.91.013846,ogawa2016classical}.
The result of the optical intensity measurement is represented as
\begin{align}
&\bracket{\hat{a}\sub{p}^\dag(s\sub{i})\hat{a}\sub{p}(s\sub{i})}\no\\
&=\bracketii{\alpha;\beta}{\hat{U}\sub{c}\hat{a}\sub{p}^\dag(s\sub{i})\hat{a}\sub{p}(s\sub{i})\hat{U}\sub{c}^\dag}{\alpha;\beta}\no\\
&=\left\|\hat{U}\sub{c}\hat{a}\sub{p}(s\sub{i})\hat{U}\sub{c}^\dag\ket{\alpha;\beta}\right\|^2\no\\
&=\left\|\int\dd s_1\int\dd s_2 f^*(s_1,s_2;s\sub{i})\hat{a}_1(s_1)\hat{a}_2(s_2)\ket{\alpha;\beta}\right\|^2\no\\
&=\left\|\int\dd s_1\int\dd s_2 f^*(s_1,s_2;s\sub{i})\alpha\psi\sub{f1}(s_1)\beta\psi\sub{f2}(s_2)\ket{\alpha;\beta}\right\|^2\no\\
&\propto
\left|\int\dd s_1\int\dd s_2 f^*(s_1,s_2;s\sub{i})\psi\sub{f1}(s_1)\psi\sub{f2}(s_2)\right|^2,
\end{align}
where we use the relation $\hat{a}_i(s)\ket{\alpha;\beta}=\alpha\psi_{\mathrm{f}i}(s)\ket{\alpha;\beta}$.
We note that the final expression above is proportional to Eq.~(\ref{eq:10}).
Therefore, the measured intensity distribution in this system exhibits the same pattern as the time-forward system.

More generally, the initial and final states can be mixed states.
The detection probability in the case of mixed initial and/or final states can be reconstructed by summing up the detection probabilities in the case of pure initial and/or final states with appropriate weighting factors.
Therefore, we can reconstruct the same detection patterns as the time-forward system including mixed states by using the classical time-reversed system employing pure initial and final states, and summing up these intensity patterns with appropriate weighting factors \cite{kaltenbaek2008quantum,PhysRevA.91.013846}.

\section{Detailed calculations of time-reversed systems}

We provide the detailed calculation of the time-reversed two-photon Young interferometer in Sec.~\ref{sec:young}, and that of the time-reversed two-photon beam focusing system in Sec.~\ref{sec:spot}. 

\subsection{Time-reversed two-photon Young interferometer}\label{sec:young}

We consider the time-reversed two-photon Young interferometer shown in Fig.~\ref{fig:2}(b).
When the source's lateral position is $x_0$, its electric field is represented as $E_0(x)\propto\delta(x-x_0)$.
The beam propagates inversely in the time-reversed system.
Initially, the first 2-$f$ system transforms the electric field into
 \begin{align}
E_1(x)&\propto\mathcal{F}[E_0(x')]\left(\frac{2\pi x}{f\lambda}\right)
\propto\exp\left(\frac{-\ii 2\pi x_0x}{f\lambda}\right),
\end{align}
where $\mathcal{F}[g(x')](k):=\int\dd x'g(x')\ee^{-\ii x'k}$ denotes the Fourier transform.
Beyond the 2-$f$ system, the beam passes through the double slit, and the electric field is transformed into
\begin{align}
E_2(x)&\propto\exp\left(\frac{-\ii 2\pi x_0 x_1}{f\lambda}\right)\delta(x-x_1)\no\\
&\hspace{1cm}+\exp\left(\frac{\ii 2\pi x_0x_1}{f\lambda}\right)\delta(x+x_1).
\end{align}
Next, the beam undergoes an optical Fourier transform (here, we assume that the light propagates in free space over a long distance $L_1$), and then the beam passes through the second 2-$f$ system.
In total, the beam is transmitted through a magnifying optical system with magnification factor $-L_1/f$, and the electric field becomes
\begin{align}
E_3(x)&\propto E_2\left(-\frac{L_1}{f}x\right)\no\\
&\propto
\exp\left(\frac{-\ii 2\pi x_0x_1}{f\lambda}\right)\delta\left(x+\frac{x_1f}{L_1}\right)\no\\
&\hspace{1cm}+\exp\left(\frac{\ii 2\pi x_0x_1}{f\lambda}\right)\delta\left(x-\frac{x_1f}{L_1}\right).
\end{align}
The beam is focused onto the nonlinear crystal for second-harmonic generation (SHG); consequently, the fundamental electric field is up-converted into 
\begin{align}
E_4(x)&\propto E_3(x)^2\no\\
&\propto
\exp\left(\frac{-\ii 4\pi x_0x_1}{f\lambda}\right)\delta\left(x+\frac{x_1f}{L_1}\right)\no\\
&\hspace{1cm}+\exp\left(\frac{\ii 4\pi x_0x_1}{f\lambda}\right)\delta\left(x-\frac{x_1f}{L_1}\right).
\end{align}
After SHG, the second harmonic (SH) beam propagates in free space over a long distance $L_2$, and the electric field is Fourier-transformed into
\begin{align}
E_5(x)&\propto\mathcal{F}[E_4(x')]\left(\frac{4\pi x}{L_2\lambda}\right)\no\\
&\propto\cos\left(\frac{4\pi x_0x_1}{f\lambda}-\frac{4\pi fx_1x}{L_1L_2\lambda}\right).
\end{align}
Finally, the SH beam is filtered by a pinhole at $x=0$.
The subsequent measured intensity is given by 
\begin{align}
I(x_0)|_{x=0}\propto|E_5(0)|^2\propto\frac{1}{2}\left[1+\cos\left(\frac{8\pi x_1x_0}{f\lambda}\right)\right],
\end{align}
which is equivalent to Eq.~(\ref{eq:11}).
Therefore, we note that the time-reversed two-photon Young interferometer exhibits the same interference pattern as the time-forward system.

\subsection{Time-reversed two-photon beam focusing system}\label{sec:spot}

We next consider the time-reversed two-photon beam focusing system shown in Fig.~\ref{fig:5}(b).
When the point light source's lateral position is $\vct{r}_0$ and its axial position from the focal plane of the first (right) 2-$f$ system is $z_0$, its lateral distribution of the electric field is given by $E_0(\vct{r})\propto\delta^{(2)}(\vct{r}-\vct{r}_0)$.
The initial 2-$f$ system transforms the electric field into
\begin{align}
E_1(\vct{r})
\propto (f+z_0)\exp\left(\frac{-\ii\pi z_0|\vct{r}|^2}{f^2\lambda}\right)\exp\left(\frac{-\ii 2\pi\vct{r}_0\cdot\vct{r}}{f\lambda}\right).
\end{align}
The electric field that is filtered by a circular aperture with diameter $D$ is further transformed into
\begin{align}
E_2(\vct{r})
&\propto\mathrm{circ}\left(\frac{\vct{r}}{D}\right)
(f+z_0)\exp\left(\frac{-\ii\pi z_0|\vct{r}|^2}{f^2\lambda}\right)\no\\
&\hspace{3cm}\times\exp\left(\frac{-\ii 2\pi\vct{r}_0\cdot\vct{r}}{f\lambda}\right),
\end{align}
where
\begin{align}
\mathrm{circ}(\vct{r}):=
\begin{cases}
1\quad (|\vct{r}|\leq 1/2)\\
0\quad (|\vct{r}|> 1/2)
\end{cases}
\end{align}
represents an aperture function. 
Next, the beam undergoes an optical Fourier transform (free-space propagation over a long distance $L_1$), and then the beam passes through the second 2-$f$ system. 
In total, the beam is transmitted through a magnifying optical system with magnification factor $-L_1/f$, and the electric field is transformed into
\begin{align}
&E_3(\vct{r})
\propto E_2\left(-\frac{L_1}{f}\vct{r}\right)\no\\
&\propto \mathrm{circ}\left(\frac{L_1\vct{r}}{fD}\right)
(f+z_0)\exp\left(\frac{-\ii\pi z_0L_1^2|\vct{r}|^2}{f^4\lambda}\right)\no\\
&\hspace{3cm}\times\exp\left(\frac{-\ii 2\pi L_1\vct{r}_0\cdot\vct{r}}{f^2\lambda}\right).
\end{align}
The beam is focused onto the nonlinear crystal for SHG; consequently, the fundamental electric field is up-converted into
\begin{align}
&E_4(\vct{r})
\propto E_3(\vct{r})^2\no\\
&\propto \mathrm{circ}\left(\frac{L_1\vct{r}}{fD}\right)
(f+z_0)^2\exp\left(\frac{-\ii 2\pi z_0L_1^2|\vct{r}|^2}{f^4\lambda}\right)\no\\
&\hspace{3cm}\times\exp\left(\frac{-\ii 4\pi L_1\vct{r}_0\cdot\vct{r}}{f^2\lambda}\right).
\end{align}
After SHG, the SH beam propagates in free space over a long distance $L_2$ and then the electric field is Fourier transformed into
\begin{align}
&E_5(\vct{r})
\propto \mathcal{F}[E_4(\vct{r}')]\left(\frac{2\pi\vct{r}}{L_2\lambda}\right)\no\\
&\propto 
(f+z_0)^2
\int_{|\vct{r}'|\leq\frac{fD}{2L_1}}\dd\vct{r}'
\exp\left(\frac{-\ii 2\pi z_0L_1^2|\vct{r}'|^2}{f^4\lambda}\right)\no\\
&\hspace{0.5cm}\times\exp\left(\frac{-\ii 4\pi L_1\vct{r}_0\cdot\vct{r}'}{f^2\lambda}\right)\exp\left(\frac{-\ii 2\pi\vct{r}\cdot\vct{r}'}{L_2\lambda}\right).
\end{align}
Finally, the SH light is filtered by a pinhole at $\vct{r}=0$.
The subsequent measured intensity is given by
\begin{align}
I&(\vct{r}_0,z_0)|_{\vct{r}=0}
\propto|E_5(0)|^2\no\\
&\propto
\bigg|(f+z_0)^2
\int_{|\vct{r}'|\leq\frac{fD}{2L_1}}\dd\vct{r}'
\exp\left(\frac{-\ii 2\pi z_0L_1^2|\vct{r}'|^2}{f^4\lambda}\right)\no\\
&\hspace{3cm}\times\exp\left(\frac{-\ii 4\pi L_1\vct{r}_0\cdot\vct{r}'}{f^2\lambda}\right)\bigg|^2.
\end{align}
After the variable transformation $\vct{r}'\rightarrow\frac{f}{L_1}\vct{r}'$, this intensity distribution exhibits the same form as Eq.~(\ref{eq:7}).
Therefore, the time-reversed two-photon beam focusing system exhibits the same intensity distribution as the time-forward system.


\begin{thebibliography}{30}
\bibitem[{foo()}]{footnote2}
\bibinfo{note}{The operation corresponding to the double slit is represented
  not by a unitary operator but the projection operator
  $\sum_{n=0}^\infty(\ket{n,x_1}\bra{n,x_1}+\ket{n,-x_1}\bra{n,-x_1})$, where
  $\ket{n,x}:=\hat{a}^\dag(x)^n\ket{0}/\sqrt{n!}$. The overall time evolution
  before two-photon detection cannot be regarded as unitary evolution, and
  therefore, this system does not correspond to be the case assumed in
  Sec.~\ref{sec:time-reversal-method}. This problem can be avoided by regarding
  the operation before the double slit as unitary evolution $\hat{U}$, and the
  operation after the double slit (including the double slit) as projection
  onto final state $\ket{\ff}$. The same is true for the case of the two-photon
  beam focusing system.}
\end{thebibliography}


\end{document}